# Slotted ALOHA Overlay on LoRaWAN – a Distributed Synchronization Approach

Tommaso Polonelli, *Student Member, IEEE*, Davide Brunelli *Senior Member, IEEE*, and Luca Benini, *Fellow, IEEE*

*Abstract*—LoRaWAN is one of the most promising standards for IoT applications. Nevertheless, the high density of end-devices expected for each gateway, the absence of an effective synchronization scheme between gateway and end-devices, challenge the scalability of these networks. In this article, we propose to regulate the communication of LoRaWAN networks using a Slotted-ALOHA (S-ALOHA) instead of the classic ALOHA approach used by LoRa. The implementation is an overlay on top of the standard LoRaWAN; thus no modification in pre-existing LoRaWAN firmware and libraries is necessary. Our method is based on a novel distributed synchronization service that is suitable for low-cost IoT end-nodes. S-ALOHA supported by our synchronization service significantly improves the performance of traditional LoRaWAN networks regarding packet loss rate and network throughput.

## I. INTRODUCTION

The Internet of Things (IoT) vision is gaining popularity, but it is not a secret that scalability, with billions of pervasive and ubiquitous devices that can potentially communicate with users or among them (machine-to-machine), is becoming a major concern because of the high density of connections and the high bandwidth desired for data transfer. At the same time, the power budget of wireless IoT communication remains tightly constrained, for the autonomy of battery-operated portable "things".

In this scenario, LoRa (Long Range) modulation, proposed by Semtech [1], is one of the most promising wide-area IoT technologies due to adaptive data rate modulation technology. It permits long-range communication at a low cost of components and reduced power consumption, affordable for battery operated devices. LoRa is supported by LoRa Alliance that has defined the LoRaWAN networking protocol, standardizing the higher-layer protocols on top of the physical radio to regulate secure communication. One potential scalability issue in current LoRaWAN is the adoption of ALOHA, a pure random-access MAC (Medium Acces Control) protocol where devices transmit "at will" without any carrier sensing. This has a detrimental effect to LoRaWAN downstream (data collection) network bandwidth, as highlighted by the simulation study reported in [2], which confirmed that increasing the number of gateways can improve the global performance but cannot eliminate fast saturation.

In fact, ALOHA is well known for its bandwidth limit when the number of nodes increases. Nevertheless, the reason for the revival of such a simple protocol is dual: (i) curtailing transceiver power and protocol complexity by avoiding carrier sensing; and (ii) no assumption on distributed synchronization among network nodes, as needed for time-based MAC, because IoT end nodes are expected to be extremely cheap (few € or less), and low-cost oscillators used in these devices drift very fast [3].

In this paper, we demonstrate that precise synchronization can be achieved even with low-cost components. Thus, LoRaWAN networks can be enhanced to support a slotted ALOHA overlay thereby achieving a significant increase in network downlink bandwidth at zero extra cost in hardware and with no changes to the standard LoRaWAN firmware. The contribution of the paper is twofold: (i) we introduce a time synchronization service for low-cost IoT devices connected to a gateway that uses the LoRaWAN protocol. It is implemented as a service at the application server layer, and as a function call at the end-node side; thus, synchronization is transparent to any version of LoRaWAN stack used; (ii) we demonstrate that regulating the access to the medium within slots (S-ALOHA) can be implemented and outperforms the standard ALOHA MAC in real-life deployments. This policy is implemented on the end-node side without any change to the LoRaWAN stack and libraries compiled for specific microarchitectures, and furthermore, it can be used with any release of the protocol; even those already installed in legacy deployments.

The rest of the article is organized as follows: after the related works, Section III presents the LoRaWAN and S-ALOHA improvements. Section IV describes the synchronization strategy, while Section V discusses the experimental results. Lastly, Section VI concludes the paper with comments and final remarks.

## II. RELATED WORKS

Large-scale IoT installations are becoming a reality, as networks are being deployed for smart city, intelligent transportation systems, urban monitoring applications. Several radio technologies (Sigfox, LoRa™, IEEE 802.15.4, NB-IoT, BLE 5.0) are currently competing in the arena of Device-to-Device (D2D) long-distance, low-power communication [4]. LoRa is a strong contender in this field, and it is actively studied. Tests done in [5][6][7] portray LoRa applications regarding network throughput and power consumption. The works in [8][9] characterize the performance of the radio link in industrial environments, while the contribution in [10] shows experiments on the coverage of LoRa. The analysis of LoRa network capacity are presented in [11], that proposes LoRa-Blink, a protocol to support multi-hop communications. Some recent works, [10]-[12], provide evaluations of LoRa link behaviour in open spaces. Unfortunately, LoRaWAN scalability is not deeply investigated in the literature, and to the best of our knowledge, no previous work proposes full solutions to address the scalability challenge of LoRa networks.

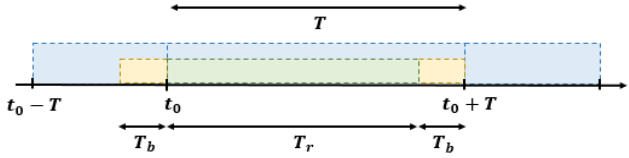

Fig. 1. Slot width definition

### III. NETWORKING MODELS

The LoRaWAN network protocol differentiates between end-devices and gateways. Three classes of end-devices are defined: Class A, Class B, and Class C.

Whilst End-devices of Class B and Class C are designed for end nodes with no critical energy constraints, and synchronized beacons are available by specification; Class A is designed for battery-powered nodes that have to be ultra-low power with very long expected lifetime. It permits bi-directional communications, but downlink transmissions are constrained to two short intervals after each uplink transmission to the gateway (Fig. 3). The MAC of LoRaWAN is based on Pure ALOHA (P-ALOHA).

As well known, an asynchronous ALOHA protocol could generate substantial inefficiencies when the number of nodes in a network increase significantly. When a message collision happens, packets are re-transmitted later, because there is no carrier check before transmitting and no listen-before-talk strategy. The presence of collisions and the need of retransmitting the packet after a collision decrease the capacity of the communication channel. The maximum ALOHA channel throughput is 18% of the overall channel capacity. Slotted ALOHA (S-ALOHA) protocol has been widely used in local wireless communications [13]. In S-ALOHA systems, the channel time is divided into slots (Fig. 1), and each terminal is enabled to transmit packets only at the beginning of a slot. Each slot with length T, reference in Fig. 1, is composed of two parts: transmission time ($T_r$) and the confidence interval ($T_b$).

If two or more terminals transmit their packet at the same time, a collision occurs and no data are successfully transmitted; otherwise no collision is generated, and the data are properly sent. In IDLE slots no terminal accesses in the channel and no device can start the transmission in the middle of the time-slot. The maximum S-ALOHA channel throughput is 37%.

The software overlay presented in this paper works on top of LoRaWAN MAC and force (Fig. 2) all the stack layers to operate into an S-ALOHA time slot. The proposed solution can be integrated into almost every application and LoRa node, since no external component or demanding computational resource is required.

Due to local rules imposed by ISO/IEC ISM regulations, wireless sensor devices working on ALOHA MAC access cannot occupy more than 1% of the channel time.

### IV. SYNCHRONIZATION

#### A. RTC synchronization on top of LoRaWAN

A synchronization procedure is fundamental to define the slots used by S-ALOHA. We implemented a lightweight synchronization library for low-cost MCU enabled devices and tested it successfully on ARM Cortex M4 microarchitecture. The end-device uses an inexpensive external crystal at 32.768 KHz with a precision ranging from 20 to 80 ppm/°C. We use this low-cost clock to generate a time reference to identify ALOHA slots and keep them aligned in all the end-devices. However, such a low-cost time-keeping component drifts significantly with time; hence we need to re-synchronize end-node clocks frequently to keep slots sufficiently aligned. To implement a synchronization scheme, we exploited the predefined answer (ACK) of a LoRaWAN Class A packet that is received with a maximum error of ± 20 μS delay [1].

To develop a robust synchronization protocol, common time "events" for both devices (sensor node and gateway) must be found to be used as a useful common reference for the synchronization. Both node and gateway must know when to open a receive/transmit window (node/gateway) after a valid packet, and the LoRa Specification describes in detail how to perform a connection ACK [1]. A timestamp on the packet sent/received (node/gateway) is saved with the goal to open the ACK windows at the same time; thus it is possible to use this event as a basis for the clock synchronization procedure.

Our method implements the scheme described in Figure 4: (1) When a packet is transmitted, the gateway and the node save the timestamp of "end of transmission". The flight time (light speed) is considered negligible; (2) Gateway considers its timestamp as reference and sends it with the ACK in the RX1 windows; (3) The node receives the new timestamp and calculates the time difference between $TX_{timestamp}$ and $RX_{timestamp}$ from which it obtains the delay offset; (4) The node adds the offset to the timestamp and updates its real-time clock; (5) From this point, both time references are synchronized.

We performed extensive measurements on the synchronization procedure, to characterize timing uncertainty due to variable execution time of function calls, transmission delays, and node-to-node variability. We found a worst-case uncertainty of 15ms, with an average of 10ms. Hence we take 15ms as a lower bound for Tb. Note that such an uncertainty is not small in absolute terms, for example in [14] and [15] synchronization uncertainties below 1ms are reached. However, our procedure works on top of an unmodified LoRaWAN stack, both as node firmware and Gateway / Server structures, instead of using physical layer modifications ([14] and [15]), hence, this level of timing accuracy is quite promising.

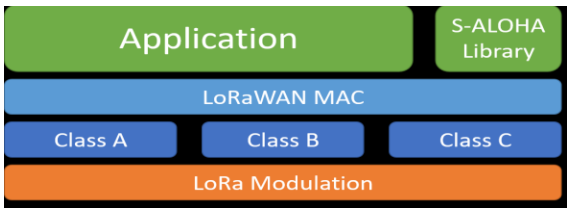

Fig. 2. LoRaWAN Stack with S-ALOHA library implemented on top of LoRaWAN MAC

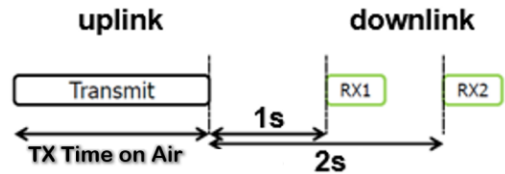

Fig. 3. LoRaWAN Uplink

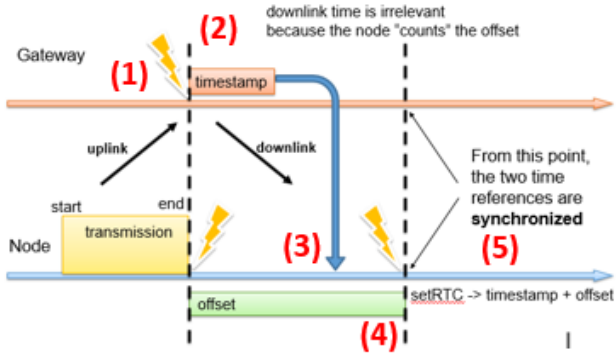

Fig. 4. Synchronization procedure

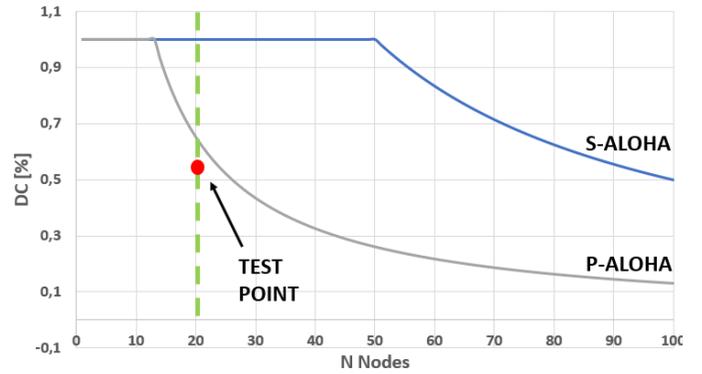

Fig. 5. Maximum Duty Cycle (DC) allowed in comparison with N nodes and channel access methodology

### B. Automatic RTC synchronization

The synchronization algorithm must guarantee an adequate end-node RTC re-alignment rate with the reference time clock for the Gateway, without a significant communication overhead and remaining compliant with LoRaWAN specifications. This is possible using the LoRaWAN Uplink and the method described above. When the packet is received from the Gateway, a timestamp is saved both on node and Gateway (Fig. 4), and an RX window is already scheduled after 1 second [1] (ACK confirmation). If the timestamp saved on Gateway is sent back with the ACK packet, the sensor node MCU will have all the information needed for clock re-alignment. Overhead is very low, only 8 bytes of timestamp, with good scalability and configurability. In many cases, there is no overhead: for example, in a heavy traffic network when end-nodes generate many data packets, re-synchronization comes for free, piggybacked on all the ACKs. When nodes need to communicate very sporadically with the gateway, some additional synchronization phases will be needed to ensure clocks do not drift too much. Maximum tolerable clock drift depends on S-ALOHA packed duration, whose sizing is application specific, and it is discussed in the next section.

## V. EXPERIMENTAL RESULTS

A LoRaWAN testbed has been developed and deployed in real-life conditions to verify the overall functionality and the network throughput advantages achievable by the S-ALOHA overlay.

### A. Slot width sizing.

The goal of this experiment is to pinpoint the minimum slot duration for the S-ALOHA overlay, taking into consideration the packet transmission time, the LoRaWAN protocol time overhead and an extra padding interval $T_b$, which is needed to compensate the clock drift and the initial synchronization uncertainty described in the previous section.

LoRa modulation allows a large number of configurations, in term of spreading factor, bandwidth and coding rate in addition to payload length that can significantly modify the packet's time-on-air and the transmission range. In this paper, we selected settings that are suitable for most applications, where the transmission range or the bit rate are not a restrictive requirement. The radio packet selected consists of 200 bytes of payload and six symbols of preamble, the spreading factor is 9 with a bandwidth of 250 KHz that generates a time of air of about 546 ms. Payload length is close to the maximum allowed (255 bytes) to decrease the ratio between data and packet overhead.

Shrinking the slot width close to the packet on-air time, the network throughput should increase due to growing number of bytes sent. One hard limit to slot with shrinkage is given by LoRaWAN uplink methodology, which needs one second of delay between transmission and receiving window (Fig. 3). This means that, considering the 8 bytes added into ACK needed for synchronization procedure, the lower bound on S-ALOHA slot duration is 1.6 seconds. However, we need to add the padding interval to this lower bound. Fig. 6 represents the maximum synchronization uncertainty between nodes, and it provides fundamental information to calculate the $T_b$ period as well as the RTC refresh rate.

In the worst case, where the device integrates an 80ppm crystal, an error of 200ms is generated every 40 minutes, a period that is often too short for applications where the nodes transmit few times for day. Therefore, considering an initial synchronization uncertainty of 15ms, we selected a Tb of 400ms. This choice allows 80 minutes between two RTC refreshes and it still has an acceptable ratio Tb (400ms)/Tr (1.6s) of 25%. Thus, a 2 seconds slot width (T) for our S-ALOHA LoRaWAN is selected.

### B. Network Setup

A final network experimental deployment was carried out with 20 sensor nodes. The goal was to verify in a realistic operating condition the S-ALOHA improvements. To compare the developed LoRaWAN S-ALOHA with an existing standard LoRaWAN installation, the modulation and the payload length were selected accordingly to the pre-existing LoRaWAN setup: the radio packet consisted of 101 bytes of payload and 6 symbols of preamble, the spreading factor was 7 with a bandwidth of 125 KHz that generates an air-time of about 167 ms. Each end-node was configured to send a sensor data packet every 30 seconds. To monitor and manage the WSN (Wireless Sensor Network), node status information was collected, such as power consumption and channel conflicts. The latter is used to analyse the WSN traffic and issues and is presented here as a result of the previous analyses. Packet time and transmission frequency were selected to generate a Duty Cycle (DC) of 0.56% for every device. This choice is not casual because is near to the maximum throughput allowed in an ALOHA baseline network, which is approximatively 0.65% in a WSN with 20 nodes.

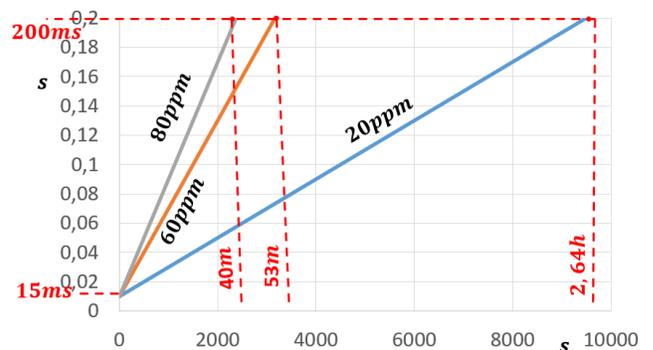

Fig. 6. Typical crystal drift over time

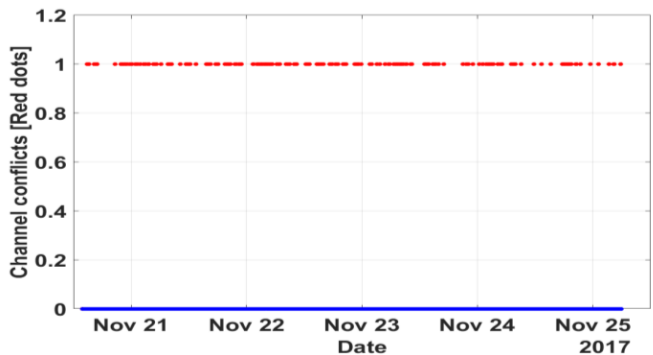

Fig. 7. P-ALOHA: Example off channel conflicts (0 → success, 1 → conflicts)

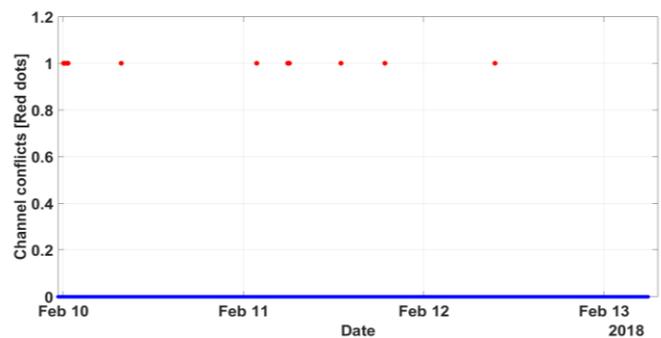

Fig. 8. S-ALOHA: Example off channel conflicts (0 → success, 1 → conflicts)

The DC presented in this paper symbolizes the maximum channel time allowed for every wireless node. Thereby, a significant difference, in term of channel conflicts is expected when comparing P-ALOA and S-ALOHA networks: the end-nodes working on the operating point do not generate enough traffic to stress the S-ALOHA network deployment. In Fig. 5, we show the maximum DC allowed as a function of N nodes and channel access methodology. The maximum value (1%) is derived from ISO/IEC ISM European regulations. Note that the horizontal curve denotes that the maximum DC is limited by regulation and does not characterize the real channel throughput. Indeed, projecting the S-ALOHA curve up to test point allows verification of real S-ALOHA improvement, which is about four times better compared to P-ALOHA. So, in other words, keeping constant the maximum DC proposed above and the packet length, a P-ALOHA LoRaWAN network can work with only 20 sensor nodes, while our S-ALOHA method can sustain more than 90 devices.

### C. Network Results

We compared the channel access collisions of P-ALOHA vs. S-ALOHA within our deployment. Tests lasted several days, and many thousands of points were acquired. Fig. 7 presents an example of test results for the P-ALOHA network. In Fig. 7, 19.927 transmissions are acquired with 301 conflicts. The overall dataset (not entirely showed in plots) for P-ALOHA, with 20 sensor nodes, consists of 188.120 transmissions with an average 1,84% probability of collision. An identical configuration was evaluated with an S-ALOHA WSN using all the methods and algorithms described above. Fig. 8 presents exemplary results for the S-ALOHA network. Note that in the initial phase the synchronization algorithm did not complete the first round for all the nodes and a cluster of collisions due to this initial transient lack of synchronization is visible on the left side of the plot. This is obviously a transient effect at start-up (or when the network is reconfigured). In Fig. 8, 18.574 transmissions are acquired with 70 conflicts. The overall dataset, with 20 sensor nodes, consists of 360.040 transmissions with an average 0,53% probability of collision, which is, as expected, more than three times lower than the one achieved by the P-ALOHA network.

## VI. CONCLUSION

We have presented the implementation of a Slotted-ALOHA overlay on top of a LoRaWAN standard protocol. S-ALOHA leverages a novel synchronization approach suitable for-low cost IoT devices and gives a 2x network throughput improvement. We demonstrate a reduction of packet collisions of 3.4x in a real-life deployment with 20 nodes operating for weeks.

ACKNOWLEDGMENT

The research contribution presented in this paper has been funded by a research grant of ST Microelectronics.